# Title: Two-particle time-domain interferometry in the Fractional Quantum Hall Effect regime

**Authors:**

I. Taktak[1], M. Kapfer[1], J. Nath [1], P. Roulleau[1], M. Acciai[2], J. Splettstoesser[2], I. Farrer[3], D. A. Ritchie[4] and D.C. Glattli[1]*,

**Affiliations:**

[1] SPEC, CEA, CNRS, Université Paris-Saclay, CEA Saclay, 91191 Gif sur Yvette Cedex France

[2]Department of Microtechnology and Nanoscience - MC2, Chalmers University of Technology, S-412 96 Göteborg, Sweden.

[3]Department of Electronic and Electrical Engineering, University of Sheffield, Mappin Street, S1 3JD, UK

[4]Cavendish Laboratory, University of Cambridge, J.J. Thomson Avenue, Cambridge CB3 0HE, UK.

**Abstract:**

Quasi-particles are elementary excitations of condensed matter quantum phases. Demonstrating that they keep quantum coherence while propagating is a fundamental issue for their manipulation for quantum information tasks. Here, we consider anyons, the fractionally charged quasi-particles of the Fractional Quantum Hall Effect occurring in two-dimensional electronic conductors in high magnetic fields. They obey anyonic statistics, intermediate between fermionic and bosonic. Surprisingly, anyons show large quantum coherence when transmitted through the localized states of electronic Fabry-Pérot interferometers, but almost no quantum interference when transmitted via the propagating states of Mach-Zehnder interferometers. Here, using a novel interferometric approach, we demonstrate that anyons do keep quantum coherence while propagating. Performing two-particle time-domain interference measurements sensitive to the two-particle Hanbury Brown Twiss phase, we find 53% and 60% visibilities for anyons with charges e/5 and e/3. Our results give a positive message for the challenge of performing controlled quantum coherent braiding of anyons.

**Introduction:**

The Quantum Hall Effect appears in high perpendicular magnetic field for electrons confined to a plane. The quantization of cyclotron orbits leads to the formation of Landau levels. For integer or fractional filling of the Landau levels, a topological insulator with a gap forms. Chiral gapless modes appear on the conductor edges on which the carriers propagate allowing a current to flow. For integer filling factor, the quantum coherence of edge channels is large. For fractional filling, the carriers are anyons whose quantum coherence is puzzling [1]: good coherence is observed on Fabry-Pérot interferometers [2-6] while non-existent [7-9] or weak [10,11] interference visibility is found in Mach-Zender interferometers, see the review [12]. In this work we use a novel kind of interferometry, based on two-quasiparticle Hanbury Brown Twiss (HBT) interference, which shows high quantum coherence of propagating anyons.

Fabry-Pérot Interferometers (FPI), based on quantum dots or antidots, showed quasi-particle interference in Fractional Quantum Hall Effect (FQHE) regime as early as the 1990s through the periodic oscillations of the transmitted current versus magnetic flux or gate voltage [3]. Recently, discrete phase shifts of the interference pattern of a FPI have been reliably ascribed to the statistical angle of anyons trapped in the dot, providing, together with an independent noise experiment [13], definitive proof of anyonic statistics [2]. In an electronic Fabry-Pérot interferometer, two separate Quantum Point Contacts (QPCs) form beam-splitters which connect a Quantum Dot (QD) to right and left leads. By appropriate tuning of their transmission, the paths of carriers going straight through the two QPCs, or performing several turns inside the dot, interfere before exiting. The interference leads to the periodic oscillation of the transmission versus the magnetic flux or versus the gate voltage used to change the dot size, attesting to the quantum coherence. The resonant character of the transmission yields quasiparticle states localized in the dot with quasi-discrete spectrum. The separation between energy levels is believed to help preserve the quantum coherence needed to observe interference.

The Mach-Zehnder Interferometer (MZI) is a different kind of interferometer, also made using two beam-splitters. Its realization in electronic systems in the Integer Quantum Hall Effect (IQHE) regime has been a breakthrough as the chiral edge channel propagation imposes a topology requiring delicate fabrication [14]. In contrast to FPIs, only two distinct paths interfere in a MZI. MZIs have been used to quantify the quantum coherence of carriers propagating along edge channels of the IQHE, in particular the loss of coherence due to the noisy environment [14-15]. Surprisingly and contrasting with measurements using FPIs, a full disappearance of interference was observed [7-9] when entering the FQHE regime (filling factor ½<$\nu$<1) and only very weak interference was observed on ultra-short MZI for filling 1/3 [12,11]. One possible reason for the different behavior could be ascribed to the nature of states involved in the two interferometers: discrete versus continuous spectrum, the latter being more fragile with respect to environmental fluctuations [16]. Counter-propagating neutral modes are also believed to degrade the coherence. Fundamental anyonic phase fluctuations may also impact the MZI visibility. The puzzling MZI visibility requires searching for quantum interferences using a different kind of interferometer, also based on delocalized propagating states.

This is the aim of this work. Here, we use a single beam-splitter to perform time-controlled quantum interference of propagating quasi-particles in the IQHE and FQHE regime. While FPI and MZI interferometers, based on DC conductance measurements, are sensitive to single-quasi-particle interference, our measurements detect current fluctuations (quantum shot noise) which is known to reveal two-particle interference, see [17,18,19]. The experiment is inspired from the seminal work of Rychkoff, Polianski, and Büttiker [20] who proposed a new kind of interferometry which is sensitive to the measurement of the so-called Hanbury Brown Twiss (HBT) quantum phase. They showed that, when applying two phase-shifted ac potentials of equal magnitude and frequency on two different contacts of a four probe conductor, the current noise reveals a two-particle interference resulting from particle indistinguishability. Here, we use the simplest four-terminal conductor: a QPC which mixes and partitions two incoming chiral edge channels into two transmitted and reflected edge channels. Fig. 1a shows the principle of the two-particle time domain interferometry measurement and Fig. 1b is a sketch of the experimental set-up used. The experiments are first done in the IQHE regime at filling factor $\nu$=2, as a benchmark situation, and then in the FQHE regime at filling factor ν=2/5 allowing us to probe anyons with charges e*=e/5 and e/3. Note that, the ν=2/5 FQHE state maps to the ν=2 IQHE state in the composite fermion picture [21] as both filling factors have two co-propagating edge channels.

Two AC sinewave voltages $V(t) = V_{ac}\cos(2\pi t)$ and $V(t-\tau) = V_{ac}\cos(2\pi t - \Delta\Phi)$ are applied to contacts (1) and (2) respectively to inject photo-created electron-hole pairs in the beam-splitter input leads, see Fig. 1a. $\Delta\Phi = 2\pi f \tau$ is the relative phase-shift due the time-delay τ between the sources. The scattering amplitudes relating input leads (1) and (2) to output leads (3) and (4) are $s_{3,1}=s_{4,2}=t$ and $s_{3,2}=s_{4,1}=ir$ (to make expressions simpler, in the main text we disregard in the scattering amplitudes the phase factors $e^{i\varphi_{\beta\alpha}}$, where $\varphi_{\beta\alpha}$ is the phase accumulated by electrons propagating from input contact (α) to output contacts (β)). $|t|^2 = D$ and $|r|^2 = 1 - D$ are the transmission and reflection probabilities of the QPC beam-splitter. Electron and hole interferences are detected through the cross-correlated current fluctuations of leads (3) and (4). According to [20], the magnitude of the current cross-correlation is shown to depend on the Hanbury Brown Twiss phase $\chi = arg(s_{13}^* s_{32} s_{24}^* s_{41})$ resulting from the indistinguishability of photo-created electron-hole pairs. Contrasting with pure DC transport and noise experiments, the creation of photo-assisted electron-hole pairs allows us to probe the HBT phase, providing a novel interferometry to test the quantum coherence of carriers. This is done by varying the time delay between the AC sources and measuring the cross-correlation noise. Note that the existence of the HBT phase is only important to probe the coherence. Its absolute value however is not relevant, being sample dependent like the geometrical phase of a MZI. The cross-correlation noise expression, in the limit of single photon excitation is, for $eV_{ac} \ll hf$, [20]:

$$S_{I_3 I_4} = -e^2 f \left(\frac{eV_{ac}}{2hf}\right)^2 \left|s_{13}^* s_{41} e^{-i\Delta\phi} + s_{23}^* s_{42}\right|^2 \quad (1)$$

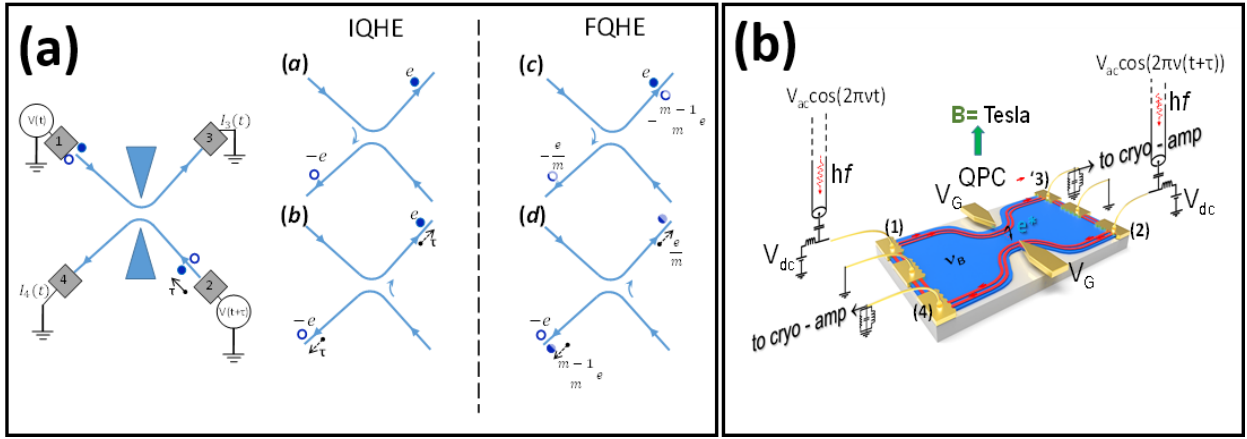

**Fig. 1 | two-particle dynamical interference principle and schematic experimental implementation. a, electron-holes pairs emitted by the AC biased contacts (1) and (2) are scattered into contacts (3) and (4) by a tunable QPC which forms an electronic beam splitter. In the IQHE regime, there are two scattering processes, where, in *(a)*, the electron-hole pair is created in lead (1) and where, in *(b)*, the electron-hole pair is created in lead (2). They both lead to an electron in (3) and a hole in (4). Indistinguishability leads to a probability for occurrence of process *(a)* and *(b)* whose variation with the relative time delay τ reveals two-particle interference. Two similar interfering scattering processes (not shown) lead to a hole in (3) and an electron in (4). This generates the cross-correlated current noise $S_{I_3 I_4}$ which is measured in the set-up sketched in Fig. 1b. In the FQHE regime, the scattering processes are similar but involve e/m fractionally charged anyons (m=3 or 5). Photo-**

**created electron-hole pairs give rise to scattering of two indistinguishable processes where in *(c)* *((d))* an electron (hole) is transmitted while a hole (electron) is backscattered as a charge $-e/m$ $(e/m)$ and transmitted as charge $-(m-1)e/m$ $((m-1)e/m)$ respectively. b. Sketch of the measurement principle. Two separate coaxial lines bring the microwave excitation on contact (1) and (2) to generate electron-hole pairs. The fluctuations of the transmitted and reflected current are converted into voltage fluctuations at contact (3) and (4). After frequency filtering and cryogenic amplification the cross-correlated noise spectrum is computed.**

Expression (1) contains the square of the sum of two two-particle probability amplitudes corresponding to the two possible electron-hole paths *(a)* and *(b)* shown in Fig. 1a whose indistinguishability is controlled by the time delay τ. The process, not shown, where an electron arrives in lead (4) and a hole in lead (3) gives similar interference and these two processes lead to current fluctuations, hence the noise given by (1).

The time-delay between AC voltages provides the knob to modulate the interference between the two paths. This plays the role of the magnetic flux or the gate voltage control used to vary the interference in FP and MZ interferometers. The complete expression, based on photo-assisted shot noise and including finite temperature and multi-photon absorption/emission of electrons and holes is:

$$S_{I_3 I_4} = -2e^2 f D(1-D) \sum_{l=+\infty}^{-\infty} l \left[ J_l \left( \frac{eV_{ac}}{hf} 2 \sin\left(\frac{\Delta\Phi'}{2}\right) \right) \right]^2 \left( \coth\left(\frac{lhf}{2k_B T}\right) - 2k_B T/lhf \right) \quad (2),$$

where $\Delta\Phi' = \Delta\Phi - \chi + \pi$. For small $V_{ac}$ and zero temperature, this expression is equivalent to equation (1).

**Results: IQHE regime.**

The measurements are performed on a two-dimensional electron gas made in high mobility GaAs/Ga(Al)As heterojunction with electron density $n_s = 1.11 \times 10^{15} m^{-2}$ and zero field mobility µ= 300 Vm$^2$s$^{-1}$. The filling factor ν=2 occurs at field B=2.3 Tesla. Low frequency conductance measurements, done while varying the QPC gate voltage $V_G$, are first performed to measure the transmitted and reflected differential conductance by applying a small (few µV) amplitude 270Hz frequency AC voltage on contact (1). The conductance traces are shown in Fig. 2a. The $e^2/h$ conductance plateau for $V_G < -0.37$ volts signals the full reflection of the ν=2 inner edge channel. We choose to partition the inner edge channel only while the outer edge channel is fully transmitted. Similar observations can potentially be obtained when fully reflecting the inner edge and partitioning the outer edge. We therefore concentrate on two working points at $V_G = -0.2V$ and -0.27V respectively corresponding to partial transmissions D=0.89 and 0.84 of the inner edge channel.

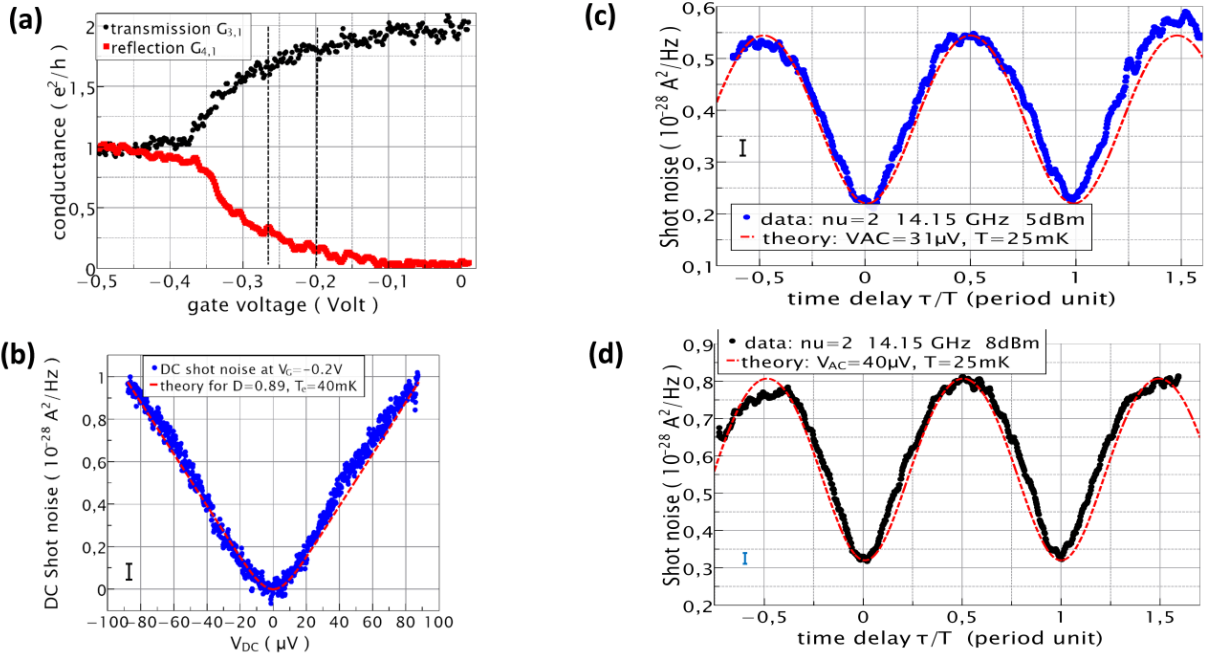

**Fig. 2 | Two-particle interferometry in the IQHE regime. a,** DC conductance trace versus gate voltage used to measure the transmission and reflection of the inner edge channel while the outer edge channel is transmitted at filling factor ν=2 . The vertical dashed lines denote the working points used here. **b,** DC Cross-correlated noise ($-S_{I_3I_4}$) data (blue circles) versus DC voltage bias $V_{DC}$ and comparison with equation (4) for transmission D=0.89 and temperature 40mK (red dashed line). The expected statistical error is $\pm 4.2\ 10^{-30}$A$^2$/Hz for 1.5s measurement time per DC voltage bias point. **c** and **d,** blue and black circles data points show the shot noise measured versus the time-delay for 14.15GHz microwave excitation and RF source powers 5 and 8 dB respectively. The dashed red curves provide comparison with the heuristic model given by equation (3) including the noise offset $S_M$. The statistical noise errors error bars are expected $\pm 3.10^{-30}$A$^2$/Hz for 3s measurement time.

Fig. 2c and 2d show two-particle interference noise measurements for a microwave frequency of 14.15 GHz at $V_G$=-0.2V for two microwave excitations differing in power by 3dB power of the sources. For the chosen frequency and 20mK temperature, the photon energy is $hf \approx 35.k_BT$ so that thermal effects are small. Clear two-particle interference is observed when varying the time delay. The visibility, calculated by the ratio of the difference between the maximum and the minimum noise over the sum of the maximum and minimum, is found to be 43% and 44% for both microwave excitations. Measurements at smaller QPC transmission, $V_G$=-0.27V, for two different AC excitations also differing in power by 3dB give similar 38 and 40% visibility, see the Supplementary Note A4.

The above measurements demonstrate two-particle quantum interference in the IQHE regime. The electron-hole pair interference can be put in perspective with electronic Hong Ou Mandel experiments where, instead of electron-hole pairs, single electrons are periodically emitted by on-
5

demand single electron source [22-25]. In Ref.[24], a driven quantum dot in the IQHE regime (also ν=2) injecting single electrons in the outer edge of a QPC beam-splitter gave the first evidence of two-particle dynamical interference. Later, Hong Ou Mandel interference with 100% visibility was observed at zero magnetic field in Ref.[25] using a different on-demand single electron source generating single particle states called levitons. In Ref.[24] electrons were injected on the outer edge only and the low visibility was attributed to the Coulomb interaction between inner and outer co-propagating edge channels giving spin-charge separation. In the present experiment, a 100% visibility may have been expected from equation (2). Indeed, as theoretically shown in [26-29], no loss of visibility is expected when including Coulomb interaction between co-propagating edge channels as the AC voltages generate coherent states which do not suffer from decoherence due to inter-channel Coulomb interaction. To understand the finite but reduced 40% visibility observed here, we suggest that this may be due to the weak elastic mixing of co-propagating channels [29] resulting from local impurities combined with spin-orbit coupling. Indeed, for different inner and outer edge channel velocities, the HBT phase for electrons injected in the inner edge and reaching the QPC in the outer edge differs from the HBT phase of electrons emitted in the outer edge channel and remaining in this channel. One can also say that the tunneling between the outer and inner edge input leads allows the inner edge channel to gain information on "which-path" the incoming electron-hole pairs are taking, thus breaking indistinguishability. In Supplementary Note A4 we give reasonable numbers supporting this, based on the channel mixing modeling of [29] briefly recalled in Supplementary Note B2. The loss of coherence can be brought in correspondence with the one observed in "which-path" experiments done with a standard MZ interferometer [33]. Further studies addressing a systematic analysis of visibility versus edge channel length should confirm the mixing hypothesis that we put forward here. We leave this issue for future works.

In order to perform a quantitative analysis, one can remark that equation (2) can be expressed as a sum of DC shot noises, i.e. shot noise without microwave excitation, where $V_{DC}$ is replaced in the expression by $lhf/e$ and weighted by the term $\left[J_l\left(\frac{eV_{ac}}{hf}2\sin\left(\frac{\Delta\Phi'}{2}\right)\right)\right]^2$

$$S_{I_3 I_4} = S_M + \sum_{l=+\infty}^{-\infty}\left[J_l\left(\frac{eV_{ac}}{hf}2\sin\left(\frac{\Delta\Phi'}{2}\right)\right)\right]^2 S_{I_3 I_4}^{DC}\left(V_{DC}=\frac{lhf}{e},T\right) \quad (3)$$

with:

$$S_{I_3 I_4}^{DC}(V_{DC},T) = -2e^2 D(1-D)eV_{DC}\left(coth\left(\frac{eV_{DC}}{2k_B T}\right) - 2k_B T/eV_{DC}\right) \quad (4)$$

In equation (3) we have added an extra noise term $S_M$ which may account for a possible mixing of co-propagating edge channels, see below. This heuristic approach provides the best fit of data. As explained in the Supplementary Note B2, it is supported by channel mixing hypothesis [29] which leads to a closed expression, see Supplementary Eq.(S6), where a mixing noise contribution is found not to depend on τ, like the heuristic term $S_M$ in equation (3). Elastic tunneling between co-propagating edge channels is likely to occur. We found a mixing tunneling probability of about 10% for our 18μm incoming channel length. This indicates a few 100μm scattering length which is compatible with typical scattering lengths ranging from few tens of μm to 100 μm reported at filling factor 2 [30-32].

To analyze the noise interference data, we use the DC shot noise measurements versus $V_{DC}$ applied on contact (1) for the gate voltage $V_G$=-0.2V shown in Fig. 2b. The red dashed line compares the data to equation (4) using the known transmission. Knowing the DC shot noise parameters extracted from the dashed red curve fit of Fig. 2b, microwave frequency and phase difference, we are left with only two unknown parameters, $V_{ac}$ and $S_M$, to analyze and fit the two-particle noise interference using the heuristic model (3).

The red dashed curves in Fig. 2c and 2d are best fits using equation (3). One finds $V_{ac}$=31μV, $S_M$=2.1 $10^{-29}$ A$^2$/Hz, and 40 μV, $S_M$=3.2 $10^{-29}$A$^2$/Hz, respectively for the two different excitations differing by a 3dB power increment. Similar fits for $V_G$=-0.27V give $V_{ac}$=33.5μV and 41 μV respectively using the heuristic formula (3). Slightly higher amplitudes are found using the complete channel mixing model of ref.[29] (see the Supplementary Note A4). For both gate voltages, the ratio of the microwave amplitudes for 3dB power increment is close to √2 , albeit slightly lower, giving confidence in the analysis. We now turn to the investigation of two-particle dynamical interference in the FQHE regime.

**Results in the FQHE regime**

We concentrate on the so-called weak backscattering regime $1 - D \ll 1$ for which the quasi-particle scattering is best understood. Consider, for simplicity, an edge channel for which the tunneling excitations carry a charge e/3. Similar reasoning can be done for charge e/5. Electrons in the edge channel form a correlated one-dimensional quantum liquid, densely occupying one state over three. To the lowest order in the backscattering amplitude, an electron can be either transmitted as a whole, charge e, with amplitude of probability $t \approx 1$, or be split as a backscattered charge e/3 and a transmitted charge 2e/3 with amplitude of probability $ir$. The same can occur for a hole, with respective charges –e, –e/3 and -2e/3 and amplitude (ir)*. Now consider an electron-hole pair created in input lead (1) and focus on the event where the hole is backscattered. After scattering, we are left with a charge e transmitted to output lead (3) and a split hole transmitted as charge -2e/3 in lead (3) and reflected as hole of charge –e/3 in lead (4), see Fig. 1c. This process leads to a total charge e/3 in lead (3) and –e/3 in lead (4). The same set of charges in the output contacts is also obtained for the process where an electron-hole pair is created in input lead (2) and the electron is backscattered. After scattering, the electron is split as a charge e/3 in lead (3) and transmitted as a charge 2e/3 in lead (4) while the hole is transmitted as charge –e in lead (4) as shown in Fig. 1d. When only charge is considered, the two events of Fig. 1c and 1d are not distinguishable and interfere.

Another process, leading to charge –e/3 and e/3 in the output leads (3) and (4) respectively, gives similar interference. The anti-correlated fluctuations of +/-e/3 charges give rise to a noise whose expression is similar to equation (1). When written to first order in reflection probability 1-D, it yields, disregarding the propagation phase accumulated in the leads:

$$S_{I_3 I_4} = -4(e^*)^2 \ f(1-D)\left(\frac{e^* V_{ac}}{2hf}\right)^2 \left|1 - e^{-i\Delta\Phi}\right|^2 \quad (5)$$

This expression agrees with the low $V_{ac}$ limit of a full multiphoton expression which has been theoretically obtained using perturbative approaches, including interactions [34-36]:

$$S_{I_3 I_4} = \sum_{l=+\infty}^{-\infty}\left[J_l\left(\frac{e^* V_{ac}}{hf} 2\sin\left(\frac{\Delta\Phi'}{2}\right)\right)\right]^2 S_{I_3 I_4}^{DC}\left(V_{DC} = \frac{lhf}{e^*}, T\right) \quad (6)$$

where:

$$S_{I_3 I_4}^{DC}(V_{DC}, T) \approx -2(e^*)^2 \ (1-D)e^* V_{DC}\left(\coth\left(\frac{e^* V_{DC}}{2k_B T}\right) - 2k_B T/e^* V_{DC}\right) \quad (7)$$

The expressions are similar to those of the integer regime Eq.(4), except for the quasi-particle charge e* and the limit 1-D<<1.

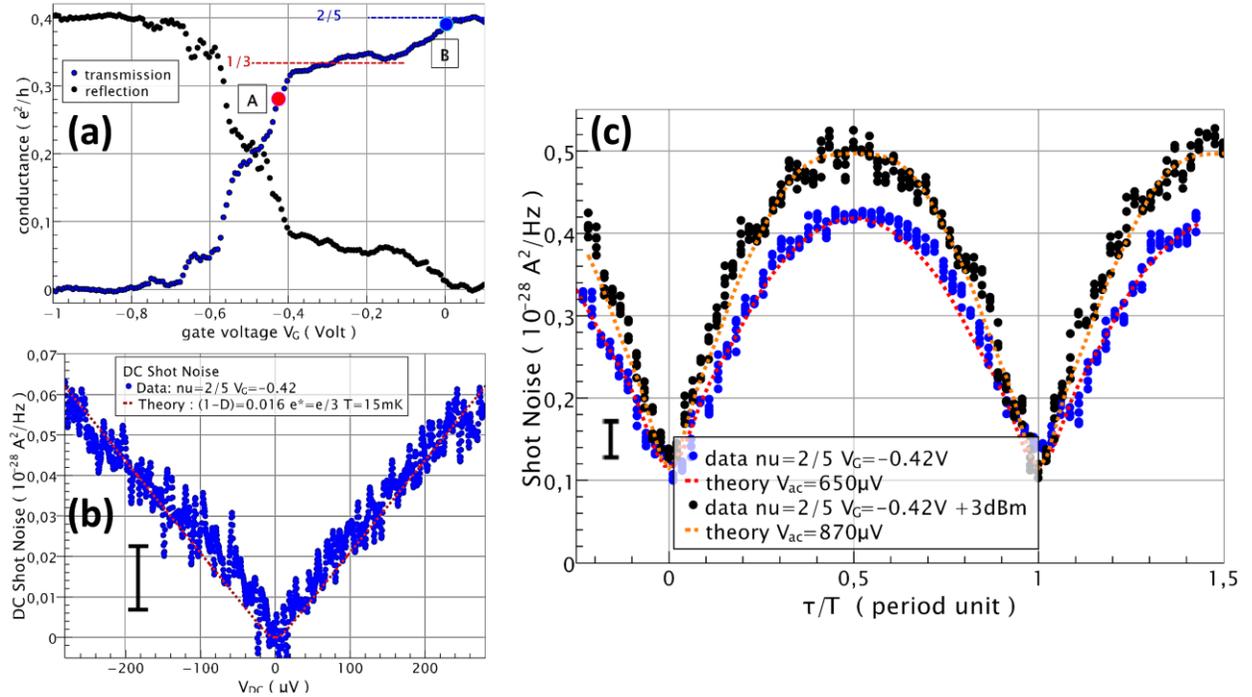

**Fig. 3 | Two-particle dynamical interference of e/3 anyons in the FQHE regime. a,** QPC conductance data versus gate voltage $V_G$ at field B=11.3Tesla ( ν=2/5 ). A plateau at conductance $\frac{e^2}{3h}$ signals the reflection of the first inner 2/5 edge channel and the formation of a 1/3 fractional state in the QPC. The red and blue filled circles, labelled A and B, indicate the working points used for probing the coherence of respectively e/3 and e/5 anyons. **b,** DC Cross-correlated noise (($-S_{I_3 I_4}$) data (blue circles), measured at working point A, and comparison with equation (7) for constant transmission D=0.984, $T_{el.}$=15mK) and e*=e/3 (red dashed curve). The statistical measurement noise errors are $\pm 0.8 \ 10^{-30} A^2/Hz$ for the typical 15s measurement time. **c,** shot noise data (black and blue filled circles) versus the time-delay τ measured at working point A for a 14.15GHz microwave excitation and two RF source powers differing by 3dB. The red dashed lines are comparisons with equation (6) using $e^* =$

$\frac{e}{3}$ including a small $10^{-29}$ A²/Hz noise offset. The statistical measurement noise errors are $\pm 2$ $10^{-30}$A²/Hz for the 3s measurement time per time delay point used.

Fig. 3a shows the conductance of the QPC for filling factor ν=2/5 in the bulk (B=11.3 Tesla). Starting at $\frac{2}{5}e^2/h$ for $V_G$=0.2V, the conductance decreases to reach a plateau at $\frac{1}{3}e^2/h$ for $V_G$< -0.1V signaling the reflection of the 2/5 inner edge channel and the formation of a region of filling factor $\nu_G$=1/3 inside the QPC. For $V_G$<-0.38V, the last edge channel starts to be reflected. We choose two working points A and B, at $V_G$=-0.42V and -0.0 Volts respectively, corresponding to the weak backscattering of anyons with charge e*=e/3 and e*=e/5 respectively. These fractional charges have been previously confirmed in a similar regime through the measurements of their Josephson relation in a previous work, see [37]. For the weak backscattering regime of the $\nu_G$=1/3 state, at working point A, Fig. 3c shows the shot noise data versus time delay for microwave powers differing by 3dB and microwave frequency f=14.15GHz. Clear oscillations with 55% and 60% visibility are observed.

To analyze the data, we introduce a constant noise offset $S_M$ to take into account a possible mixing noise, as done for the IQHE study. The fits, calculated from equation (6) with e*=e/3 and using the measured DC shot noise data taken in the same conditions, are shown as red dashed curves. They provide an excellent agreement with each experimental trace with $V_{ac}$= 650+/- 20 µV and $V_{ac}$= 870+/- 10 µV and $S_M$=0.11 $10^{-28}$ A²/Hz. For a 3dB microwave power difference, the ratio of the $V_{ac}$ values is 1.34+/-0.05, close to √2. Note that the theoretical analysis would have required DC shot noise data up to $eV_{DC} \cong 10hf$ while the range of measurements in Fig. 3b is limited to $V_{DC} \cong 7hf$. Such extrapolation has been also safely used in Ref.[37] for similar conditions.

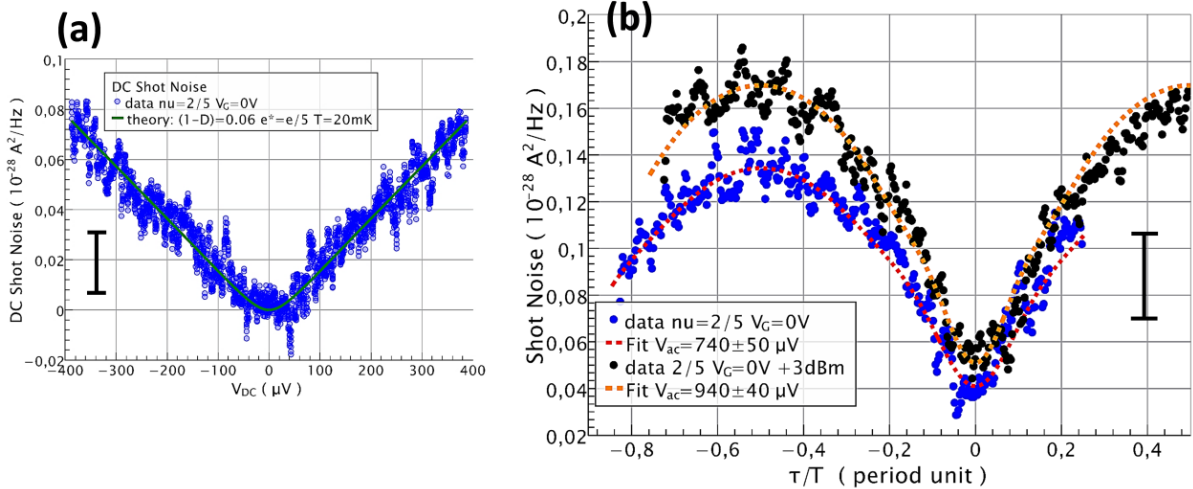

**Fig. 4 | Two-particle dynamical interference of e/5 anyons in the FQHE regime. a, DC Cross-correlated noise ($-S_{I_3I_4}$) data (blue circles), recorded at working point B. The red dashed curve is computed from equation (7) for transmission D=0.94 and $T_{el.}$=20mK (green solid curve) and e*=e/5. Statistical measurement noise errors are expected to be $\pm 11$ $10^{-30}$A²/Hz for the 10s measurement time. b, the black and blue filled circles are shot noise measurement data versus time-delay taken at working point B for 14.15GHz microwave excitation and two RF source powers differing by 3dB. The red dashed curves are computed from equation (6) using e*=e/5 and temperature 20mK including a 4 $10^{-30}$ A²/Hz noise offset. The statistical**

measurement noise errors are $\pm 0.9 \; 10^{-30} A^2/Hz$ for the 15s acquisition time per time-delay point.

Similar measurements performed for $\nu_G=2/5$, at working point B, are displayed in Fig. 4b for the same 14.15GHz frequency and two microwave powers differing by 3dB. Visibilities of 53% and 51% are observed. The fits done using e*=e/5 and experimental DC shot noise traces give $V_{ac}=$ 740+/- 50 µV and $V_{ac}=$ 940+/- 40 µV, with a $V_{ac}$ ratio 1.27+/- 0.15 weaker but still close to √2 for 3dB power difference. Here the weaker noise due to the one fifth charge leads to larger uncertainty due to lower signal to noise ratio.

The fact that a higher visibility is found at filling factor 2/5 than at filling factor 2 may be at first sight surprising. It indicates that less channel mixing occurs between co-propagating edge channels in the FQHE regime than in the IQHE regime. Note, however, that no direct comparison can be done between these two cases since the underlying mixing mechanisms are likely to be different. Possible reasons for this are the spin polarization at 11Tesla ($\nu=2/5$) which cannot be compared with that at a field of 2.3 Tesla ($\nu=2$), as well as the reduced inter-channel tunneling at low energies, which is expected from chiral Luttinger liquid physics [39]. These effects motivate further theoretical modelling.

.

To conclude, the present demonstration of two-particle dynamical interference in the FQHE regime shows that the propagating edge channels do keep significant quantum coherence over several tens of microns (our sample dimension), contrasting with the experiments performed using MZIs. The work also provides the first example of a novel interferometry based on the HBT phase. This work helps resolve the issue of diverging results previously reported on quantum coherence in the FQHE regime, at the same time highlighting the work needed on MZI to understand the prior results better. The current demonstration also highlights the possibility of performing anyon braiding in FQHE systems, which is a crucial step towards the realization of topologically protected qubits for quantum computing.

**Methods:**

Experiments are done in a Cryoconcept dry dilution refrigerator with 20mK base temperature equipped with 14.5 Tesla dry superconducting magnet. Conductance measurements are done using lock-in amplifiers at 270 Hz frequency and 2µV excitation voltage. The two microwave excitations used for interferometry are provided using two synchronized DC-40GHz room temperature sources with relative phase control ability. About 60-70dB microwave attenuation between the source and the sample is provided by cryogenic 50 Ohms attenuators. Noise measurements are made using home-made cryogenic amplifiers followed by fast digital acquisition. The FFT cross-correlation computation is done on a PC computer providing real-time noise acquisition.

**Data and materials availability:**

All data, code, and materials used in the analysis are available in some form to any researcher for purposes of reproducing or extending the analysis. The raw experimental shot noise data of all figures are available as a single zip file from [40].

**Acknowledgments:** We thank P. Jacques for technical help, P. Pari, P. Forget and M. de Combarieu for cryogenic support, and we acknowledge discussions with J. Rech, Th. Jonkheere, A. Crépieux, I. Safi, Th. Martin, Y. Gefen, M. Goldstein, A. Das, S. Manna and members of the Saclay Nanoelectronics team.

**Funding:** This work was funded by the ANR FullyQuantum 16-CE30-0015-01 grant, the H2020 FET-OPEN UltraFastNano #862683 grant, and the EPSRC grant K004077.

**Author contributions:**

D.C.G. designed and supervised the project.

I.T. performed the experiment with help from D.C.G. and J.N.;

M.K. fabricated the sample on heterojunctions grown by I.F. and D.R.

I.T., D.C.G., J.S., M.A. and P.R. analyzed and discussed the data

I.T.& D.C.G. wrote the manuscript with inputs from all coauthors;

* main corresponding author: christian.glattli@cea.fr

**competing interests:** none declared

**Supplementary Information for: "Two-particle time-domain interferometry in the Fractional Quantum Hall Effect regime"**

**Authors:**

I. Taktak[1], M. Kapfer[1], J. Nath[1], P. Roulleau[1], M. Acciai[2], J. Splettstoesser[2], I. Farrer[3], D. A. Ritchie[4] and D.C. Glattli[1*],

**Affiliations:**

[1] SPEC, CEA, CNRS, Université Paris-Saclay, CEA Saclay, 91191 Gif sur Yvette Cedex France

[2] Department of Microtechnology and Nanoscience - MC2, Chalmers University of Technology, S-412 96 Göteborg, Sweden.

[3] Department of Electronic and Electrical Engineering, University of Sheffield, Mappin Street, S1 3JD, UK

[4] Cavendish Laboratory, University of Cambridge, J.J. Thomson Avenue, Cambridge CB3 0HE, UK.

A. Experimental Methods
   1. Sample characteristics and fabrication
   2. Measurement set-up
   3. Visibility optimization
   4. IQHE measurements at $V_G$=-0.27V
B. Electron-hole pair Interferometry
   1. The Hanbury Brown Twiss phase
   2. Channel mixing modeling
   3. Comparison with MZI visibility loss due to similar channel mixing:
C. Supplementary references

**Supplementary Note A1. Sample characteristics and fabrication:** samples are 2DEGs with electrons confined at the interface of high mobility epitaxially grown GaAs/GaAlAs heterojunctions at 90 nm below the surface. The low temperature zero field mobility is $2.10^6 \text{cm}^2\text{s}^{-1}\text{V}^{-1}$ and the electron density is $n_s$=1,11 $10^{15}\text{m}^{-2}$. For this density, the bulk filling factor $\nu_B$=2/5 corresponds to a magnetic field of ≈11.2 Tesla. Ohmic contacts are realized by evaporating 125 nm Au, 60 nm Ge, 4 nm Ni followed by annealing at 470°C. A shallow mesa etching (H3PO4 phosphoric acid, time 4 minutes) defines the sample. The QPC gates are realized by e-beam lithography, see Fig.S1 for a SEM image of the sample used. The edge channel length between each contacts and the center of the QPC is about 18μm. .

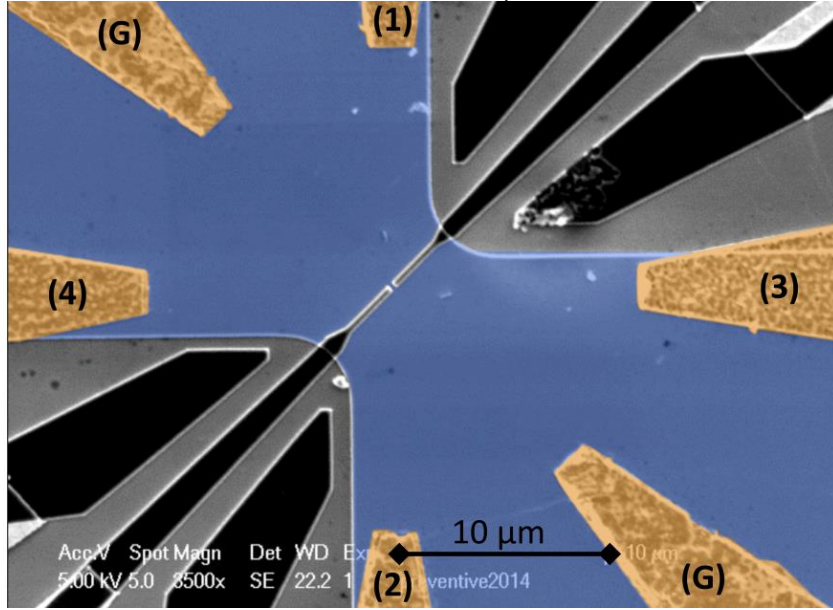

**Figure S1: SEM view of the sample used. Yellow areas are ohmic contacts, blue aeras denote the unetched part of 2DEG mesa. A 10um black bar indicates the scale.**

**Supplementary Note A2. Measurement set-up:** an ultra-low temperature cryo-free dilution refrigerator with a 20 mK base temperature from CryoConcept is used as in [1,2]. It is equipped with a dry superconducting coil able to reach 14.5 Tesla. Ultra-low-loss dc-40GHz microwave cables, same as in ref. [1,2], bring the room temperature microwave excitation from an Agilent N5183A RF source to a Printed Circuit Board (PCB). The RF power of the microwave source, given in the main text is attenuated by fixed 63dB cold attenuators and extra losses in the cryogenic coaxial cables. Coplanar waveguides designed by CST microwave Studio$^R$ etched on the PCB bring the two radiofrequency excitations to ohmic contact (1) and (2) of the sample, see Fig.1(f) and Fig.S1. Noise measurements are obtained by separately converting the transmitted and reflected current fluctuations into voltage fluctuations at contact (3) and (4) respectively in parallel to a R-L-C resonant circuit tuned to 2,2 MHz frequency and bandwidth ≈ 150 kHz, with R=20kOhms. Note that an effective resonant circuit resistance $R_{eff.} = RR_L/(R + R_L) \approx$ 6.5kOhms is found instead of 20kOhms due to inductance loss, giving a shunt resistance $R_L = (L2\pi f_0)^2/r$ in parallel to R, where r=15 Ohms is the series resistance of the inductance. Finally, the Q factor of the RLC resonant circuit is given by the ratio of the parallel resistance $R_{eff}//R_{Hall}$ to the characteristic impedance $\sqrt{\frac{L}{C}}$, where $R_{Hall}$ is the Hall resistance of the sample. The voltage fluctuations are amplified by two home-made cryogenic amplifiers with 0.22 nV/Hz$^{1/2}$ input noise at low temperature, followed by low noise room temperature amplifiers. The amplified fluctuations are passed through Chebyshev filters and then sent to a fast 20Ms/s digital acquisition card (ADLink 9826) while a PC provides real-time computation of the cross-correlation spectrum.

Absolute Noise calibration is done by recording the equilibrium Johnson Nyquist noise when varying the temperature from 20mK to 200mK. Differential Conductance measurements giving the transmission and reflection are made by applying a low frequency AC voltage, frequency 270Hz, and µV amplitude voltage to contact (1) and sending the amplified AC voltage from contacts (3) and (4) to two Lock-in amplifiers. The low frequency measurement accuracy is mostly limited by the large 1/f noise of the cryogenic HEMT (white noise cross-over at ≈1MHz). The shot noise accuracy is limited by the input white noise of the amplifier and time averaging. For $\nu_B$=2/5, the 20kOhm resistor and the ≈5kOhm inductance the effective RLC parallel resistance $R_{eff}$=6.5kOhms in parallel with the bulk Hall resistance converts the input noise of 220pV/Hz$^{1/2}$ into $1.4 \cdot 10^{-27}$A$^2$/Hz equivalent current noise power. Using cross-correlation and noise averaging during the measurement time $\tau_m$ =3s with ≈150kHz effective detection bandwidth, the accuracy of a raw noise data point is +/- 2 $10^{-30}$A$^2$/Hz.

**Supplementary Note A3. Visibility optimization:**

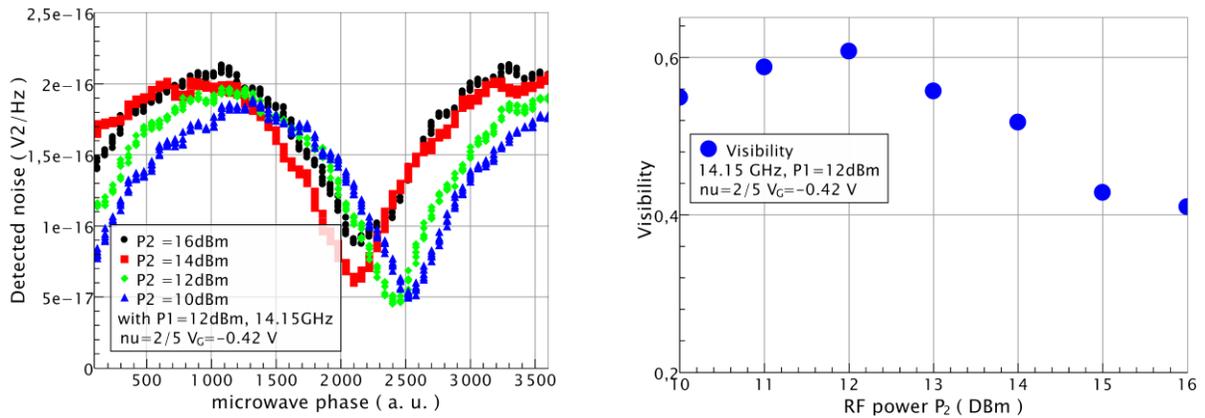

**Fig. S2: Visibility optimization.**

**The left graph shows noise measurements versus time-delay for various RF source powers P$_2$ sent to contact (2) while the power sent to contact (1) is P$_1$= 12dBm. The equality of the actual V$_{ac}$ amplitudes transmitted on contacts (1) and (2) is obtained for the maximum of the visibility. Right graph: visibility versus power P$_2$ in dBm.**

In the main text, all noise interference formulas are given for equal RF amplitudes applied on the injecting contacts. As the microwave transmission of the RF-lines is not accurately known at low temperature and may suffer from unwanted microwave reflection in connectors linking different temperature stages or suffer from uncontrolled dissipation, it is best to find a way to determine in situ the equality of the RF amplitudes. This is done by fixing the microwave amplitude on contact (1) and varying the amplitude on contact (2) while measuring the noise versus time-delay. The maximum of visibility signals equal RF amplitudes and allows to fix P2, here about -0.5dB lower than P1 to ensure perfect equality of the amplitudes on the sample.

## Supplementary Note A4. IQHE measurements at $V_G=-0.27V$

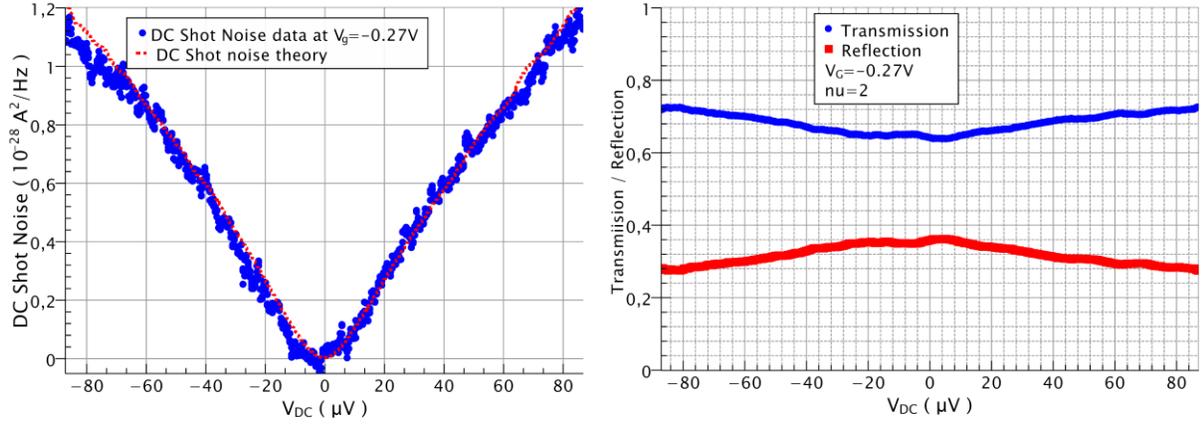

**Fig. S3: right: transmission and reflection probability for the partial transmission of the inner edge channel versus bias voltage at $V_G=-0.27V$ and filling factor 2. Left: DC shot noise measured in the same conditions. The red dashed curve is calculated using the DC shot noise theoretical expression (4) in which a weak energy dependent transmission deduced from the right graph is included.**

Here we present measurements which are complementary to the ones presented in the main text. The mean inner edge channel reflection is R=0.3 for $V_G=-0.27V$. Figure S3 shows a slight non-linearity of the transmission versus DC voltage bias. The DC shot noise trace (blue filled circles) is well described by the standard DC shot noise formula, equation (4), using the measured transmission.

Figure S4 shows the two-particle interference shot noise for 5dBm and 8dBm RF source power at frequency 14.15GHz (blue filled circles). The red dashed line is a fit using the heuristic equation (3) for a constant transmission D=0.7 and AC voltage amplitude $V_{ac}$=33.5 and 41 µV for respectively 5 and 8 dBm. The extracted values are close to the ones deduced with similar RF power at the other gate voltage -0.20V. The black dashed line is a fit using a model of channel mixing, see equation (S6) below.

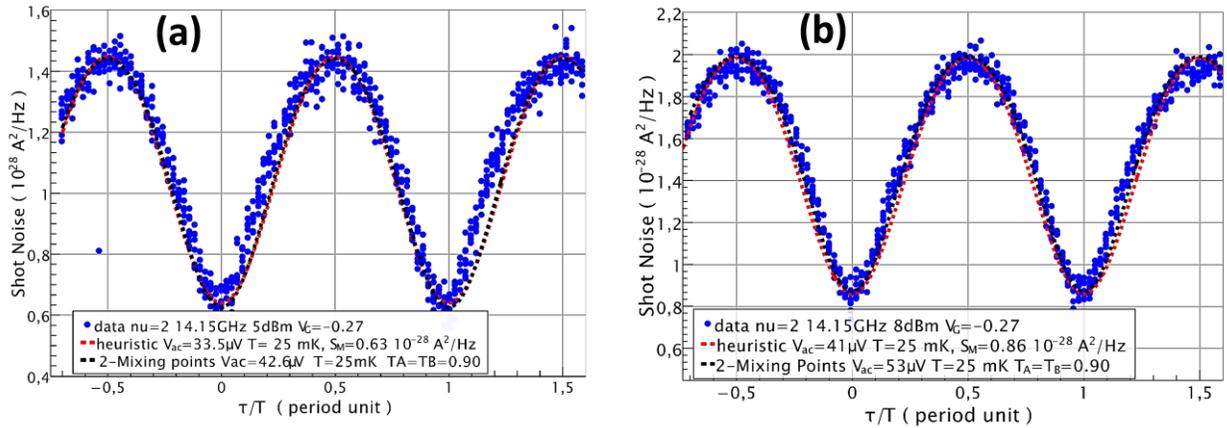

**Fig. S4:** Shot noise versus time-delay measured for 14.15GHz microwave excitation and RF source powers 5dBm (a) and 8 dBm (b) respectively at $V_G$=-0.27V (blue filled circles). The comparison with the heuristic equation (3) including a noise offset $S_M$=0.63 and 0.86 $10^{-28}$ $A^2$/Hz is shown as a red dashed curve. Comparison with equation (S6) for a model of two symmetric mixing points (strength $R_A$=$R_B$=0.10, $\Delta\tau=\tau_i-\tau_o$=44ps) placed on both input channels is shown as a black dashed curve. A constant QPC transmission 0.7 and a 25mK temperature are used to fit the data.

## Supplementary Note B1. The Hanbury Brown Twiss phase:

We consider zero temperature for simplicity and no channel mixing such that the scattering of a single channel by the QPC only occurs. Electrons emitted by the reservoir at energy $\varepsilon$ and experiencing the AC voltage in contacts (1) and (2) are put in a quantum superposition of states with energies $\varepsilon_{+/-} = \varepsilon \pm hf$. For low AC amplitude, the amplitude of probability to absorb/emit a single photon is $p_{\pm 1} = \pm \left(\frac{eV_{ac}}{2hf}\right)^2$ for contact (1) and $p_{\pm 1}e^{i\Delta\phi} = \pm \left(\frac{eV_{ac}}{2hf}\right)^2 e^{i\Delta\phi}$ for contact (2) where $\Delta\phi=2\pi f\tau$ is the voltage phase shift. Higher photon number processes are discarded for $\left(\frac{eV_{ac}}{2hf}\right)^2 \ll 1$. The many-particle state incident on the scattering region is then [3]:

$$|in\rangle = |F\rangle + \int_0^{hf} d\varepsilon\, p_1 \left(\hat{a}_1^\dagger(\varepsilon)\hat{a}_1(\varepsilon_-) + e^{i\Delta\phi}\hat{a}_2^\dagger(\varepsilon)\hat{a}_2(\varepsilon_-)\right)|F\rangle \qquad (S1)$$

where $|F\rangle = \prod_{\varepsilon<E_F,\alpha=1-4} \hat{a}_\alpha^\dagger(\varepsilon)|0\rangle_\alpha$ is the ground state formed by the filled Fermi sea of all leads and $\hat{a}_\alpha^\dagger(\varepsilon)$ the fermionic creation operator acting on the occupation states of contacts α=1,2.

Denoting $\widetilde{\hat{a}_\alpha}(\varepsilon) = \hat{a}_\alpha(\varepsilon) + p_1\hat{a}_\alpha(\varepsilon - hf) + p_{-1}\hat{a}_\alpha(\varepsilon + hf)$ the annihilation operator describing, to lowest order in single photon absorption/emission probability, the electrons emitted by lead (α) and scattered in energy by the AC potential, the operators describing the states of electrons entering in contacts (3) and (4) are:

$$\hat{b}_3(\varepsilon) = s_{31}\widetilde{\hat{a}_1}(\varepsilon) + s_{32}\widetilde{\hat{a}_2}(\varepsilon) \qquad (S2a)$$

$$\hat{b}_4(\varepsilon) = s_{41}\widetilde{\hat{a}_1}(\varepsilon) + s_{42}\widetilde{\hat{a}_2}(\varepsilon) \qquad (S2b)$$

The many-particle outgoing state is then:

$$|out\rangle = |0\rangle + \int_0^{hf} d\varepsilon\, p_1 \Big[s_{31}s_{31}^*\hat{b}_3^\dagger(\varepsilon)\hat{b}_3(\varepsilon_-) + s_{41}s_{41}^*\hat{b}_4^\dagger(\varepsilon)\hat{b}_4(\varepsilon_-) + s_{31}s_{41}^*\hat{b}_3^\dagger(\varepsilon)\hat{b}_4(\varepsilon_-) +$$
$$s_{41}s_{31}^*\hat{b}_4^\dagger(\varepsilon)\hat{b}_3(\varepsilon_-) + e^{i\Delta\phi}\Big(s_{32}s_{32}^*\hat{b}_3^\dagger(\varepsilon)\hat{b}_3(\varepsilon_-) + s_{42}s_{42}^*\hat{b}_4^\dagger(\varepsilon)\hat{b}_4(\varepsilon_-) + s_{32}s_{42}^*\hat{b}_3^\dagger(\varepsilon)\hat{b}_4(\varepsilon_-) +$$
$$s_{42}s_{32}^*\hat{b}_4^\dagger(\varepsilon)\hat{b}_3(\varepsilon_-)\Big)\Big]|0\rangle \qquad (S3)$$

The terms of the form $\hat{b}_\alpha^\dagger(\varepsilon)\hat{b}_\alpha(\varepsilon_-)$ describe neutral excitations in the output leads and do not contribute to current noise at zero frequency. The only terms contributing to the zero frequency cross-correlated fluctuations of leads (3) and (4) are of the form $\hat{b}_\alpha^\dagger(\varepsilon)\hat{b}_{\beta\neq\alpha}(\varepsilon_-)$. Namely:

$$p_1\left[s_{31}s_{41}^* + e^{i\Delta\phi}s_{32}s_{42}^*\right]\hat{b}_3^\dagger(\varepsilon)\hat{b}_4(\varepsilon_-) + p_1\left[s_{41}s_{31}^* + e^{i\Delta\phi}s_{42}s_{32}^*\right]\hat{b}_4^\dagger(\varepsilon)\hat{b}_3(\varepsilon_-) \qquad (S4)$$

The first term corresponds to the interference of electron-hole pairs separately created in lead (1) and (2) and scattered as an electron in lead (3) and a hole in lead (4), as shown in figure 1(b) and (c) of the main text. The second term describes the interference leading to an electron created in lead (4) and a hole in lead (3). These terms leads to equation (1) of the main text.

Defining $s_{31} = te^{i\varphi_{31}}$, $s_{42} = te^{i\varphi_{42}}$, $s_{41} = (ir)e^{i\varphi_{41}}$, $s_{32} = (ir)e^{i\varphi_{32}}$, with $\varphi_{\beta\alpha}$ denoting the electronic phase accumulated while propagating from contact ($\alpha$) to contact ($\beta$) and using the local transmission and reflection amplitudes t and (ir) for the QPC, we see that the interference allows us to probe the so-called HBT phase [3] $\chi = arg(s_{13}^* s_{32} s_{24}^* s_{41}) = \pi + (\varphi_{32} + \varphi_{41}) - (\varphi_{31} + \varphi_{42})$ using the controlled time-delay of the sinewave sources. Expression (S4) vanishes for $\Delta\Phi = \chi - \pi$. The phase $\chi$, not accessible to DC transport and DC noise measurement, is the relevant phase of the two-particle dynamical interferometer and plays a role similar to the phase of single particle interferometers. Note that the actual value is not important because it is sample dependent like the different arms of an MZI. Even in the original work of Ref. [3], the actual value of the phase $\chi$ was not presented as fundamental.

Let us discuss in more details the interference process. One can interpret the first term in brackets in equation (S4) in the following way: when an electron-hole pair is created in lead (1) the 2-particle probability amplitude to have an electron in lead (3) and hole in lead (4) is $p_1 s_{31} s_{41}^*$. Similarly, when an electron-hole pair is created in lead (2) the 2-particle probability amplitude to have an electron transmitted in lead (3) and hole backscattered in lead (4) is $p_1 e^{i\Delta\Phi} s_{32} s_{42}^*$. As these two processes are not distinguishable, one has to add the probability amplitude of each process, hence the term in brackets. The second term in brackets in (S4) represents a similar two-particle interference but with a hole in lead (3) and an electron in lead (4). The +/-e charge fluctuations generated in output leads by these two processes have equal probability. This leads to the noise power density given by equation (1).

One can extend the two-particle interference processes to the FQHE regime. When an electron-hole pair is created, the transmitted electron (hole) is transmitted as a whole, charge e (-e) while, in contrast to the IQHE regime, the standard backscattering is replaced by a partial backscattering of the incoming charge. One finds a split hole (electron) transmitted as a charge (m-1)/me (-(m-1)/me) and backscattered as a charge e/m (-e/m). This leads to equation (7) using the weak backscattering of the scattering amplitudes. Equation (7) quantitatively agrees, in this limit, with a more involved interacting electron theory.

## Supplementary Note B2. Channel mixing modeling:

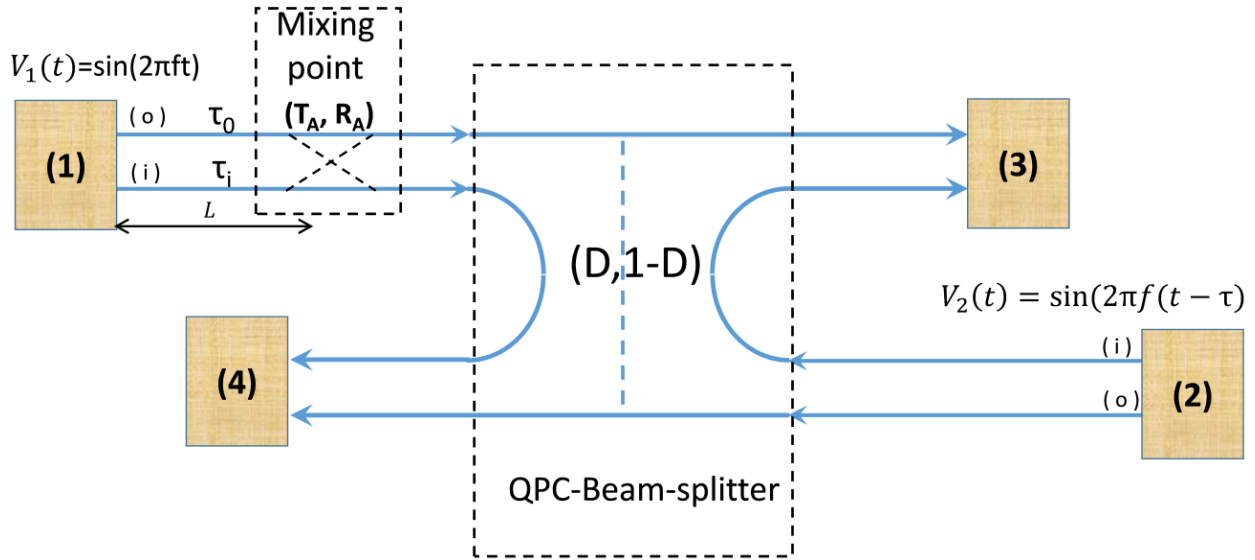

**Figure S5: elastic channel mixing.** The QPC beam splitter is aimed at partitioning the outer edge channel only, transmission D. However, a localized scatterer situated at distance L from the injecting contact (1) allows tunneling between the outer and inner edges. The probability of inter-edge tunneling is $R_A$. $T_A = 1-R_A$ is the probability of no mixing. For different outer and inner edge channel propagation velocities $V_0$ and $V_i$, electrons on the outer and inner edge have different dynamical phase $\varphi_o = 2\pi f \tau_o$ and $\varphi_i = 2\pi f \tau_i$. This generates photo-assisted shot noise making electrons arriving at the upper left of the QPC beam splitter noisy, reducing the visibility of the two-particle interference. In other words, the mixing point allows the inner edge channel to know "which path" the interfering electron-hole pairs are coming from, spoiling quantum coherence. Similar physics holds when the QPC partitions the inner edge while transmitting the outer edge.

Figure S5 schematically shows how one can model the effect of channel mixing. We consider only a single mixing point along one input edge channel (the left input). Generalization to more mixing points both on the left and right side of the central QPC beam-splitter is straightforward and gives similar qualitative effects, see Ref.[4]. Being interested in bulk filling factors $\nu=2$ and $2/5$, we consider only co-propagating edge channels.

In the absence of mixing, the inner edge channel is spectator and the noise of lead (3) or (4) measures the two-particle interference of indistinguishable electron-hole pairs photo-created in the outer at input leads (1) and (2). The visibility is expected to be 100%. However a finite mixing allows the inner edge to acquire information on which path an electron-hole pair comes from and the visibility is reduced. Note that this requires that the wave-function of particles propagating on channel (o) and (i) can be distinguished. This is indeed so if the propagation time $\tau_{i,o}$ in the inner and outer edge are different such that the dynamical phases $\varphi_{i/o}$ are different. A full calculation using standard Floquet scattering theory can be found in [4]. For a single scattering point, as considered in figure S5, one finds the shot noise $S_I$:

$$-S_{I_3 I_4}(\tau) = 2\frac{e^2}{h}hfD^2(T_A R_A)\left(\frac{eV_{ac}2\sin(2\pi f(\Delta \tau))}{hf}\right)^2 + T_A 2(e^*)^2 fD(1-D)\left(\frac{e^* V_{ac}}{2hf}\right)^2(1-\cos(2\pi f(\tau - \tau_i))) + R_A 2(e^*)^2 fD(1-D)\left(\frac{e^* V_{ac}}{2hf}\right)^2(1-\cos(2\pi f(\tau - \tau_o)))$$

(S5)

Where $\Delta\tau = \tau_o - \tau_i$ and we have assumed for simplicity that the propagation time from the scatter to the QPC is negligible. Including finite propagation time towards the QPC can be taken into account by a proper redefinition of the time delay $\tau$.

The effect of mixing is twofold: -1) the generation of an extra noise given by the first term of equation (S5) ; -2) the smearing of the interference as $\tau$ cannot make the last two terms simultaneously vanish unless $\tau_o = \tau_i = \tau$.

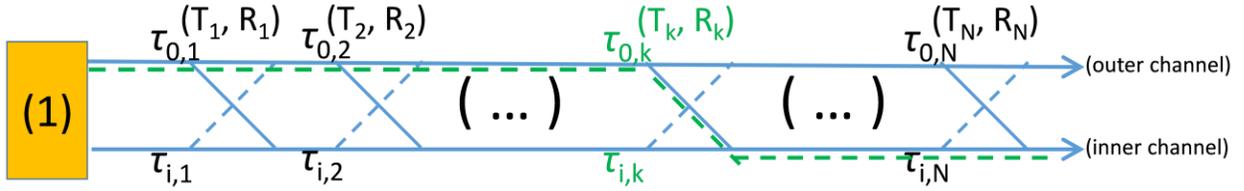

**Figure S6: multiple inter edge channel tunneling points.** $\tau_{o,k}$ and $\tau_{i,k}$ denote the propagation time between point (k) and (k-1). For weak inter-edge tunneling, $R_{k=1,N} \ll 1$, and, now, the inner channel partitioned by the QPC, one can add in the HBT interference noise the independent contribution of each tunneling point (k) weighted by the probability $R_k$ and using the time delay $\tau_k = \sum_{q \leq k} \tau_{o,q} + \sum_{q>k} \tau_{i,q}$ corresponding to the propagation along the dashed green path describing the tunneling from the outer edge to the inner edge at point (k). These contributions lead to the smearing of the HBT interference and loss of visibility.

The model can be extended to two mixing points of strength $R_A$ and $R_B$ placed on both input channels at a distance L from the injecting contacts, see [4]. Here we give the expression for symmetrically placed mixing points of equal strength. This expression is used to fit the data of figure S4 (black dashed curves).

$$-S_{I_3 I_4} = 2\frac{e^2}{h} hf \left[ D(1-D)\left( (T_A^2 + R_A^2)S(\tau) + T_A R_A (S(\tau + \Delta\tau) + S(\tau - \Delta\tau)) \right) + 2D^2 T_A R_A S(\Delta\tau) \right] \quad (S6)$$

Here, $\tau$ is the controlled time delay between the AC sources, $\Delta\tau = |\tau_i - \tau_o| = \left|\frac{L}{V_i} - \frac{L}{V_o}\right|$ is the propagation time difference for particles travelling from the contact to the mixing point with velocity $V_i$ and $V_o$ for respectively the inner and outer edge, $T_A = 1 - R_A$ and $S(\tau) = \sum_l l J_l^2 \left(\frac{e^* V_{ac}}{hf}\right) 2 \sin\left(\frac{\Delta\Phi'}{2}\right)\left( \coth\left(\frac{lhf}{2k_B T}\right) - \frac{2k_B T}{lhf} \right)$. The last term in (S6), independent of $\tau$, originates from the noise generated at each mixing point and plays the role of $S_M$ in the heuristic formula, the first three terms are interference terms. The shift by $+/-\Delta\tau$ with respect to $\tau$ in the argument of $S(\tau)$ contributes to the smearing of the HBT interference. Figure S4 uses equation (S6) to fit the experimental data. A moderate 10% mixing strength ($R_A = 0.1$) is found. This is consistent with a channel mixing length observed in the literature [5-7]. For L necessarily smaller than the 18μm input channel length, a propagation time difference of 44ps suggests an outer edge channel velocity >2.4 $10^5$m/s dominating the inner edge velocity. This is consistent with the outer and inner edge channel velocities reported for filling factor 2. A fit of the data of similar quality is obtained using two different mixing parameters $T_A = 0.92$ and $T_B = 0.86$ while keeping same $V_{ac}$ and $\Delta\tau$ values.

The smearing of the interference pattern is probably the most serious loss of visibility if multiple tunneling points occur. For example, let us consider multiple mixing points in one arm, say the left, as suggested in Fig S6, where we choose to partition the inner edge. To first order of inter-edge tunneling probabilities $\{R_p\}$, we are left with a sum of terms corresponding to tunneling paths (k) where an electron emitted from (o) has been transferred to the input edge (i) at the $k^{th}$ tunneling point corresponding to the propagation time $\tau_k = \sum_{q \leq k} \tau_{o,q} + \sum_{q>k} \tau_{i,q}$ and contributing by $R_k 2(e^*)^2 f D(1-D)\left(\frac{e^* V_{ac}}{2hf}\right)^2 (1 - \cos(2\pi f(\tau - \tau_k)))$ to the two-particle noise.

For stronger mixing $R_{k=1,N}$, second order inter-tunneling probabilities (like the case where an electron emitted in channel (i) visit once channel (o) and returns back to channel (i) to contribute to the interference) add more paths with different propagation times. This further smears the original interference pattern. In a more realistic description one has to consider multiple mixing points in both left and right input leads. This contributes to more loss of visibility.

The following graph, Fig S7, compares the visibility obtained for a single mixing point approach and that obtained by considering the continuous limit of an ensemble average of weak mixing points (multiple tunneling between outer and inner edge channel are neglected). We observe that the two curves are very similar, but in order to compare the continuous mixing point (red dashed) curve with the single mixing point (blue) curve, a slight increase (13%) of the mixing strength $R_A$ has to be done. One can conclude that, for weak mixing, a single mixing point and a continuous distribution of mixing points give equally a good qualitative and quantitative representation of the data.

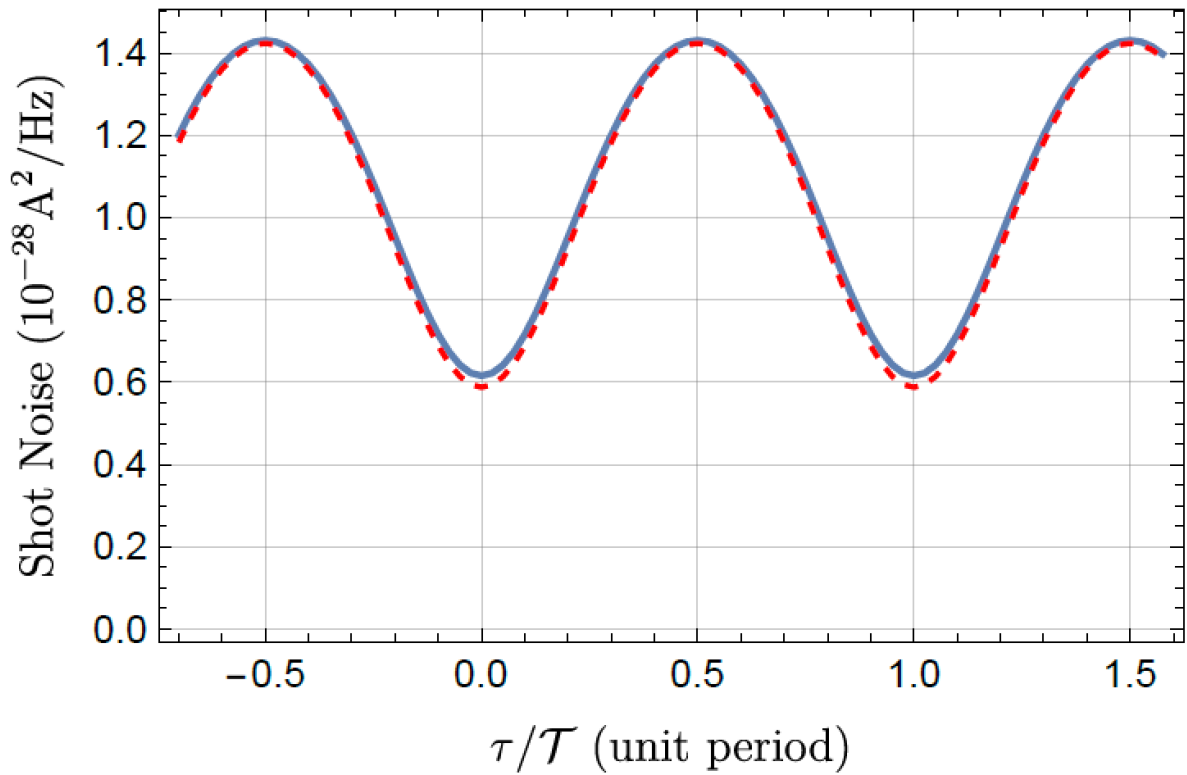

**Figure S7: Blue curve:** fit of the experimental data of Fig. S4(a) using single mixing points of strength $R_A=0.1$, $eV_{ac}/hf = 0.727$, $\Delta\tau=44$ps and 25mK temperature. The red dashed line curve is the calculation of an ensemble average of mixing points but the strength has been increased to $R_A=0.13$ while other parameters are kept constant.

## Supplementary Note B3. Comparison with MZI visibility loss due to similar channel mixing:

A direct comparison of the loss of coherence in a MZI and that observed in our experiment is not straightforward. While in a MZI, channel mixing alone is enough to induce a loss of visibility, in the present experiment having both channel mixing and different channel velocity propagation are required to explain a reduction of visibility. Nevertheless, in order to provide a tentative comparison, we present here a calculation of the MZI loss of visibility expected for a mixing strength equal to the one observed in our two-particle dynamical interference measurements.

Here we consider the effect of channel mixing on the visibility of a Mach-Zehnder interferometer. Then we will compare, for the same mixing channel, the loss of visibility for two-particle HBT interference and MZI interference.

We consider a MZI at filling factor ν=2. Interference is performed on the inner channel using two quantum point contacts of transmission $T_1$ and $T_2$. An outer edge channel is copropagating along the inner edge. In absence of inner/outer edge channel mixing, the output current $I_i$ found in the inner edge is:

$$I_i = \frac{e^2}{h} V_{DC}\left(T_1 T_2 + R_1 R_2 - 2\sqrt{T_1 T_2 R_1 R_2}\cos(\phi)\right) \tag{S7}$$

Where ϕ is the phase including a possible arm length difference and an Aharonov-Bohm flux, and the dependence in energy has been neglected for simplicity. Let us now consider a single mixing point, located at some position (A) on the upper arm and denote the tunneling probability between inner and outer edge channel as $R_A=1-T_A$. By analogy with the case of our dynamical HBT interference, we will consider that both inner and outer edges are fed by the same ohmic contact at potential $V_{DC}$ while all other contacts are at zero potential.

Let us now calculate the current $I_i$ in the inner output edge. It is made of two contributions. A first contribution to $I_i$ is an interference contribution $I_{i(i)}$ due to electrons injected in the inner edge by the polarized contact:

$$I_{i(i)} = \frac{e^2}{h} V_{DC}\left((T_1 T_A)T_2 + R_1 R_2 - 2\sqrt{(T_1 T_A)T_2 R_1 R_2}\cos(\phi)\right) \tag{S8}$$

A second contribution to $I_i$ is a non-interference contribution $I_{i(o)}$ due to electrons emitted in the outer edge by the polarized contact:

$$I_{i(o)} = \frac{e^2}{h} V_{DC}(R_A T_2) \tag{S9}$$

The total inner edge output current is:

$$I_i = \frac{e^2}{h} V_{DC}\left(R_A T_2 + (T_1 T_A)T_2 + R_1 R_2 - 2\sqrt{(T_1 T_A)T_2 R_1 R_2}\cos(\phi)\right) \tag{S10}$$

The interference visibility for a single mixing point $V_{i,1}$ is:

$$V_{i,1} = \frac{2\sqrt{(T_1 T_A)T_2 R_1 R_2}}{R_A T_2 + (T_1 T_A)T_2 + R_1 R_2} \tag{S11}$$

In the case of two impurities, one of strength $R_A$ in the upper arm and the other one of strength $R_B$ in the lower arm, we have similarly an interfering term:

$$I_{i(i)} = \frac{e^2}{h}V_{DC}\left((T_1T_A)T_2 + (R_1T_B)R_2 - 2\sqrt{(T_1T_A)T_2(R_1T_B)R_2}\cos(\phi)\right) \quad (S12)$$

And a non-interfering term similar to the previous case:

$$I_{i(o)} = \frac{e^2}{h}V_{DC}(R_AT_2) \quad (S13)$$

The interference visibility for two mixing points $V_{i,2}$ is:

$$V_{i,2} = \frac{2\sqrt{(T_1T_A)T_2(R_1T_B)R_2}}{R_AT_2 + (T_1T_A)T_2 + (R_1T_B)R_2} \quad (S14)$$

For the mixing strength $R_A=R_B = 0.1$ deduced from figure S4 the two-particle dynamical HBT interference gives a 40% visibility. For the single-particle Mach-Zehnder interference, according to the above calculation, equation (S14), a visibility of 90% is expected for the best tuning of the MZI ($T_1=T_2=0.5$).

### D. Supplementary references